\documentclass[%
 reprint,
superscriptaddress,
frontmatterverbose, 
 amsmath,amssymb,
 aps,
 pra,
]{revtex4-2}

\usepackage{lipsum}
\usepackage{xcolor}
\usepackage{graphicx}
\usepackage{dcolumn}
\usepackage{bm}
\usepackage{siunitx}
\usepackage{xcolor}
\usepackage{dsfont}
\usepackage{tikz}
\usepackage{tikz-3dplot}
\usepackage{qcircuit}
\usepackage{hyperref}
\usepackage{physics}
\usepackage{float}
\usepackage{graphicx}

\begin{document}

\preprint{APS/123-QED}


\title{Improved accuracy on noisy devices by non-unitary \\ Variational Quantum Eigensolver for chemistry applications}

\author{Francesco Benfenati}
 \altaffiliation[Present address: ]{Multiverse Computing, Donostia-San Sebasti\'an, Spain}
 \affiliation{%
Dipartimento di Scienze Fisiche e Chimiche, Universit\`a degli Studi dell'Aquila, Coppito, L'Aquila, Italy
 }

\author{Guglielmo Mazzola}
 \affiliation{%
IBM Quantum, IBM Research – Zurich, 8803 R\"uschlikon, Switzerland
 }%

\author{Chiara Capecci}
 \affiliation{%
Dipartimento di Scienze Fisiche e Chimiche, Universit\`a degli Studi dell'Aquila, Coppito, L'Aquila, Italy
 }
  \affiliation{%
Dipartimento di Fisica, Sapienza Universit\`a di Roma, Roma, Italy
 }
 
 \author{Panagiotis Kl. Barkoutsos}
 \affiliation{%
IBM Quantum, IBM Research – Zurich, 8803 R\"uschlikon, Switzerland
 }%
 
 \author{Pauline J. Ollitrault}
 \affiliation{%
IBM Quantum, IBM Research – Zurich, 8803 R\"uschlikon, Switzerland
 }%
 
  \author{Ivano Tavernelli}
  \email[]{ita@zurich.ibm.com}
  \affiliation{%
IBM Quantum, IBM Research – Zurich, 8803 R\"uschlikon, Switzerland
 }%

 \author{Leonardo Guidoni}
 \email[]{leonardo.guidoni@univaq.it}
 \affiliation{%
Dipartimento di Scienze Fisiche e Chimiche, Universit\`a degli Studi dell'Aquila, Coppito, L'Aquila, Italy
 }
 
\date{\today}

\begin{abstract}

We propose a modification of the Variational Quantum Eigensolver algorithm for electronic structure optimization using quantum computers, named non-unitary Variational Quantum Eigensolver (nu-VQE), in which a non-unitary operator is 
combined with the original system Hamiltonian leading to a new variational problem with a simplified wavefunction Ansatz.
In the present work, we use, as non-unitary operator, the Jastrow factor, inspired from classical Quantum Monte Carlo techniques for simulation of strongly correlated electrons.
The method is applied to prototypical molecular Hamiltonians for which we obtain accurate ground state energies with shallower circuits, at the cost of an increased number of measurements.
Finally, we also show that this method achieves an important error mitigation effect that drastically improves the quality of the results for VQE optimizations on today's noisy quantum computers. The absolute error in the calculated energy within our scheme is one order of magnitude smaller than the corresponding result using traditional VQE methods, with the same circuit depth. 

\end{abstract}

\maketitle

\section{Introduction}
Among the important problems that quantum computers promise to solve in the future, computational chemistry will probably be one of the first in which quantum computation might show an advantage over classical computers \cite{McArdle_QCC, Bauer2020, 2009.12472, QC}. Simulating quantum systems by manipulating quantum objects, a possibility that Richard Feynman envisioned more than 30 years ago, is becoming a reality\cite{challenges_opportunities, NISQ_technology}.
In the last years, much progress has been done in the development of computational chemistry algorithms on quantum computers. Using current quantum computer devices has been possible to experiment 
this new computational paradigm 
on small molecules, providing a proof-of-principle of the actual applicability of the quantum computing algorithms~\cite{Kandala2017HardwareefficientVQ, hempel2018quantum,colless2018computation,kandala2019error}, 
such as the Variational Quantum Eigensolver (VQE) for the optimization of the system wavefunction. 
In this introduction, we will cover only some technical limitations of this approach, which the present work aims to overcome, while we refer to recent reviews for a comprehensive treatment of the subject\cite{sim_elec_struct,McArdle_QCC,age_QC}.

The ultimate goal is to solve the Schr\"odinger equation for a system of interacting electrons within a molecule, 
defined by the Hamiltonian

\begin{equation}
  \hat  H=\sum_{p,q}h_{pq} \hat a_p^\dagger \hat a_q +\frac{1}{2}\sum_{p,q,r,s}h_{pqrs} \hat a_p^\dagger \hat a_q^\dagger \hat a_r \hat a_s
\label{H_second_quant}
\end{equation}
where $\hat a_p$ and $ \hat a_p^\dagger$ represent the annihilation and creation operators of molecular spin-orbital $p$, respectively, while $h_{pq}$ and $h_{pqrs}$ are hamiltonian parameters, which values define the particular molecule of interest in a given basis set of one-electron functions.
Next step consists in translating a quantum chemistry problem into qubit operators~\cite{kassal2011simulating}. This mapping from fermion to qubits can be achieved by using different transformations, like Jordan-Wigner \cite{Jordan1928}, Bravyi-Kitaev \cite{Bravyi2002,BK_transformation}, or parity  \cite{bravyi2017tapering}. The transformed  qubit Hamiltonian can be written as a linear combination of Pauli matrix strings~\cite{whitfield2011simulation, Moll2018, barkoutsos2018quantum}
\begin{equation}
    H= \sum_jh_jP_j=\sum_jh_j\bigotimes_i \sigma_i^j
\label{eq:qubit_hamiltonian}
\end{equation}
In Eq.~\eqref{eq:qubit_hamiltonian} the collective index $j$ labels the different Pauli strings, $P_j$ (a tensor product of Pauli matrices, $\sigma_i \in \{\sigma_x=X,\sigma_y=Y,\sigma_z=Z, \mathbb{I}_2 \}$), associated to the terms in eq.~\eqref{H_second_quant} after applying the fermion-to-qubit map, while the index $i$ runs over the qubits, each one associated to an occupied or virtual orbital of the basis set.  
For instance, in the Jordan-Wigner scheme, the state of the $i$-th qubit represents the occupation number of the $i$-th molecular spin-orbital.

 
The VQE algorithm aims to variationally approximate the ground state of the system, and has been already applied to electronic structure problems and beyond \cite{Peruzzo,Kandala2017HardwareefficientVQ,VQE_McClean_2016,Quantum_Chemistry_QPE,Quantum_Chemistry_dynamics}.
This mixed quantum-classical algorithm can approximate the ground state and the lowest energy of the Hamiltonian through the Rayleigh-Ritz variational principle. Given the choice of a variational wavefunction, which is represented by a quantum circuit that depends on some external variational parameters, the expectation value of the qubit Hamiltonian is evaluated and minimized by optimizing them through a classical minimization procedure. 
The calculation of the expectation value, i.e. the weighted sum of Pauli strings is performed by repeated measurement cycles, each time re-evaluating the same circuit upon the state collapse occurring after each measurement. 
Once the desired statistical precision in the sampling of the energy expectation value is achieved, an external optimizer updates the variational parameters and the procedure is repeated till a minimum is reached. The computational cost of VQE has a $\mathcal O(N_b^4)$ scaling in the number of spin-orbital $N_b$, that is in the same order as the number of qubits.

In the quest for a compact parametrization of the wavefunction ansatz, i.e. the circuit guess, well-established techniques used in classical quantum chemistry are an excellent source of inspiration. In particular the Unitary Coupled Cluster (UCC) technique \cite{doi:10.1002/qua.21198, UCC} provides a systematic way to construct a variational circuit adding several excitation operators. A complementary approach relies on heuristic or hardware-efficient ans\"{a}tze, \cite{Moll2018,Ganzhorn} that simply make use of the native quantum gates directly available on the quantum hardware. 
While physically motivated circuits efficiently target the relevant parts of the Hilbert space, heuristic ones usually lead to shallower circuit depths. This last characteristic is crucial due to the limited coherence times of present quantum devices.
Overall, the search for wavefunction ans\"{a}tze that are at the same time compact and accurate is still ongoing.

In this paper, we propose a method to improve the accuracy of a variational ansatz without increasing the depth of the corresponding quantum circuit. In our scheme, the increased state representability is obtained by introducing non-unitary operators in the ansatz.
In turn, these can be evaluated at the price of measuring additional Pauli strings. 
This method can be classified as a hybrid ansatz for the VQE, since a part of the complexity of the wavefunction is transferred to the target operator.
Other mixed quantum-classical methods can be found in Refs.
\cite{VQAs, OO_UCC, IBM_OO_UCC, PhysRevLett.123.130501, non_orthogonal_VQE, decoherence_mitigation}. 
This method is similar in spirit to the one presented by some of us in a previous paper, \cite{PhysRevLett.123.130501} but here we devise a different implementation, as detailed in the next paragraph.
At variance with Ref.~\cite{PhysRevLett.123.130501}, the proposed approach does not require the doubling of the qubit resources. The non-unitary Variational Quantum Eigensolver (nu-VQE)  method presented in this paper is also more general and can be applied to calculate the lowest eigenvalue of any Hamiltonian. Finally, we have observed that the use of nu-VQE has an important error mitigation effect that improves the accuracy of the VQE result by one order of magnitude under conditions relevant for present devices.

The paper is organized as follows. In Section \ref{nu-VQE method} we describe the general nu-VQE method. We then define the Jastrow operator in Section \ref{Jastrow operator} and in Section \ref{Computational_details} the computational details of the simulations are given. Finally in Section \ref{Results} we show the results obtained from exact noiseless simulations as well as from measurement-based quantum computing simulations using a noise model based on the calibration parameters of IBM's quantum computer \textit{ibmq\_boeblingen}. The conclusions follow.

\section{nu-VQE method}
\label{nu-VQE method}
In this section, we introduce a class of ans\"{a}tze featuring the product of a non-unitary operator $\hat O(\vec{\lambda})$ and a more standard unitary operator $ \hat U(\vec{\theta})$ acting on an initial state $\ket{\Psi_0}$. This non-normalized wavefunction can be written as:
\begin{equation}
 |\Psi^O(\vec{\theta}) \rangle= \hat O(\vec{\lambda})  \hat  U(\vec{\theta})\ket{\Psi_0} =   \hat  O(\vec{\lambda}) |\Psi(\vec{\theta}) \rangle
 \label{non-norm_wf}
 \end{equation}
The estimated energy of the state is given by
\begin{equation} 
E =\frac{\langle \Psi^O(\vec{\theta})| \hat H|\Psi^O(\vec{\theta})\rangle} {\langle \Psi^O(\vec{\theta})|\Psi^O(\vec{\theta})\rangle},
\label{normalization}
\end{equation}
which can be rewritten as
 \begin{equation} 
 E =\frac{\langle\Psi(\vec{\theta})| \hat O^\dagger(\vec{\lambda})  \hat H  \hat O(\vec{\lambda})|\Psi(\vec{\theta})\rangle}{\langle\Psi(\vec{\theta})| \hat  O^\dagger(\vec{\lambda})  \hat  O(\vec{\lambda})|\Psi(\vec{\theta})\rangle}.
 \label{en_eval_O}
 \end{equation}
The energy of Eq.~\ref{en_eval_O} can therefore be obtained from the expectation values of the modified Hamiltonian $ \hat O^\dagger  \hat H  \hat O$, and the operator $ \hat O^\dagger \hat  O$. The number of the additional measurements depends on the form of $ \hat O$.

The optimization of the two sets ($\vec{\lambda}$ and $\vec{\theta}$) of variational parameters can be performed with standard methods. 
With this procedure, we extended the VQE variational flexibility to a new class of non-normalized wavefunctions without increasing the circuit depth, at the price of performing additional measurements.
We note that, while the present work features a specific class of non-unitary operators other choices are also possible, provided that the operator takes the form of product of Pauli's matrices.

\section{Jastrow operator}
\label{Jastrow operator}
In this work we consider a particular choice of the operator $ \hat O$ which is inspired by the single-body and two-body Jastrow factors \cite{PhysRev.98.1479,Zen2014,Barborini2016} in Quantum Monte Carlo, similarly to the operator used in \cite{PhysRevLett.123.130501}. In classical calculations, the Jastrow factors are corrective terms that enable the construction of accurate explicitly-correlated wavefunctions, beyond the mean-field level. 
Different functional forms of Jastrow factors, with different accuracy and computational costs, have been devised in the literature.
Here we choose to construct the Jastrow operator already at the qubit level, i.e. after that  a fermion-to- qubit mapping has been applied to the physical problem. 
Moreover we include single-qubit and two-qubit operators only. This restriction allows to control the number of additional parameters to optimize.
Indeed, with $N$ the number of qubits, our Jastrow operator in matrix representation takes the following form:
\begin{equation} \label{eq:Jtot} J=J_1+J_2 \end{equation}
with
\begin{equation} \label{eq:J1} J_1=\exp\left[{-\sum_{i=1}^N {\alpha_i Z_i}}\right] \end{equation}
and
\begin{equation} \label{eq:J2} J_2=\exp\left[-\sum_{i<j=1}^N{\lambda_{i,j}Z_i Z_j}\right]\end{equation} where $\alpha_i$ and $\lambda_{i,j}$ are independent real coefficients and $Z_i=\otimes_j (\mathbb{I}_2)_{j\neq i} (Z)_{j=i}. $ 

In the form defined by Eqs.~\eqref{eq:Jtot},~\eqref{eq:J1} and~\eqref{eq:J2}, the number of additional terms to be measured grows exponentially with the number of qubits compared to the normal VQE.
However, we can replace the exponential operator with its linearized approximated form:
\begin{equation} \label{eq:J} J(\vec{\alpha},\vec{\lambda})=1-\sum_{i=1}^N{\alpha_i Z_i -}\sum_{i<j=1}^N{\lambda_{i,j}Z_i Z_j}\end{equation}
so that the number of additional terms to measure only grows with the fourth power of $N$. 
Indeed, since the Jastrow operator is applied both to the ket and the bra, one needs to measure up to 4-qubit Pauli matrix strings.

\begin{figure*}
\includegraphics[width=1.0\linewidth]{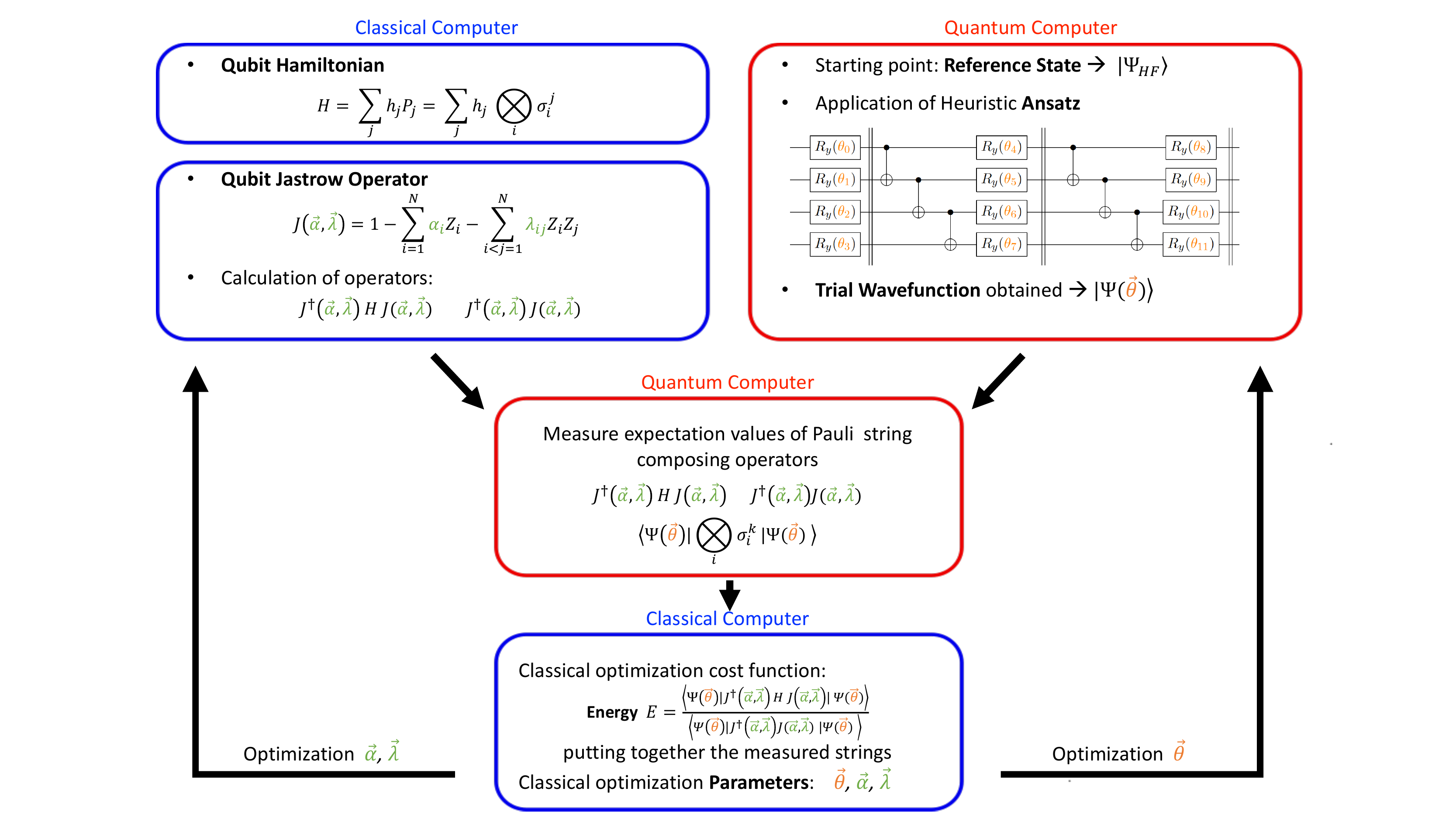}
\caption{Schematic representation of the nu-VQE algorithm presented. The blocks enclosed by blue lines summarize the operations performed by the classical computer, whereas the blocks enclosed by red lines summarize the operations performed by the quantum computer. 
}
\label{fig:procedure}
\end{figure*}

The procedure to optimize the modified Hamiltonian is represented in Figure \ref{fig:procedure} and works as follows. 

\begin{enumerate}

\item Given a set of circuits, initialize the wavefunction on the quantum computer following the usual heuristic or hardware-efficient ansatz. \cite{Kandala2017HardwareefficientVQ}

\item Given a set of parameters that define the Jastrow operator $ \hat J$, measure all the Pauli strings composing $ \hat J^\dagger  \hat H  \hat J$ and $ \hat J^\dagger  \hat J$

\item Calculate the expression 
 \begin{equation} 
 E =\frac{\langle\Psi(\vec{\theta})|\hat{J}^\dagger(\vec{\alpha},\vec{\lambda}) \hat{H} \hat{J}(\vec{\alpha},\vec{\lambda})|\Psi(\vec{\theta})\rangle}{\langle\Psi(\vec{\theta})|\hat{J}^\dagger(\vec{\alpha},\vec{\lambda})  \hat{J}(\vec{\alpha},\vec{\lambda})|\Psi(\vec{\theta})\rangle}.
 \label{en_eval}
 \end{equation}

\item Optimize all the parameters together ($\vec{\theta}$, $\vec{\alpha}$ and $\vec{\lambda}$) to minimize the energy, like in normal VQE.

\end{enumerate}

Note that $ \hat J= \hat J^\dagger$ because the Pauli matrices are Hermitian and the Hermitian adjoint is anti-linear, and it is distributive over the tensor product.
Albeit this method increases the number of variational parameters to be optimized, it dramatically reduces the the circuit depth necessary to achieve chemical accuracy.

\section{Computational details}
\label{Computational_details}

We calculate the one- and two-body integrals of the fermionic Hamiltonian using the PyQuante Python package. 
We choose either the STO-3G basis set \cite{Pople-STO} or the 6-31G basis set \cite{Pople-631g}, leading to different number of orbitals (and eventually different number of qubits).
For the generation of the qubit Hamiltonian we base our implementation on the Qiskit Python package \cite{Qiskit}. 
For the fermion-to-qubit mappings we use the Jordan-Wigner \cite{Jordan1928}, Bravyi-Kitaev \cite{Bravyi2002}, or parity transformations \cite{bravyi2017tapering}. 
For the latter transformation we also introduce symmetry reduction techniques, like the two qubit reduction, thanks to the conservation of electron number and spin \cite{bravyi2017tapering}.

The wavefunction ansatz used in this work falls within the hardware efficient class proposed by Kandala et al. in Ref. \cite{Kandala2017HardwareefficientVQ}.
The initial state, the Hartree-Fock state, is evolved under the action of the circuit unitary $U(\vec{\theta})$ to give the trial wave function $\ket{\Psi(\vec{\theta})}$.

The circuit is made of a series of blocks built from single-qubit rotations (along the y-axis), followed by an entangler $U_{\text{ENT}}$, that spans the required length of the qubit register.
In our tests, we chose the simplest choice of a ladder of CNOTs gate with linear connectivity, such that qubit $q_i$ is the target of qubit $q_{i-1}$ and the control of qubit $q_{i+1}$, with $i=1,\cdots,N-2$\cite{Moll2018, barkoutsos2018quantum}.
For a representation of the ansatz, see Figure \ref{fig:procedure}. 
All the simulations in this paper are run using 2 entangling blocks, except those in Figure \ref{fig:vs_N_Blocks}(a) where we study how the results change with different numbers of blocks. 
 The initial rotation angles are sampled from a uniform distribution from $0$ to $2\pi$, and the initial Jastrow parameters are sampled from a uniform distribution within the interval $(-0.1,0.1)$. 

We perform two different types of simulations, which we call (1) exact noiseless simulations and (2) measurement-based simulations. 
In the first case, we manipulate the so-called \emph{state-vector}, using linear algebra matrix-vector operations.
We perform these simulations using the Qiskit \cite{Qiskit} statevector simulator and QuTiP Python package \cite{QuTip}. 
Notice that in this circuit emulation we always have access to the full wavefunction (in the form of a statevector), hence we can compute the expectation values exactly as shown in the Eq. \ref{en_eval}. 
In actual devices we cannot access the full wavefunction in a scalable way, but the information about a quantum states can be accessed indirectly through measurements. 
In particular the estimation of the energy is performed as the sum of the expectation values of single Pauli operators, multiplied by the respective Hamiltonian coefficient. 
Each expectation value can be obtained by sampling from the prepared ansatz using $N$ measurements, hence $N$ repetition of the same circuit, generally called ``shots'' \cite{IBM2020_energy_calculations}.
Since the result on an actual quantum computer is inherently statistical, the energy measurement is affected by statistical error.

The second type of simulations is therefore a simulation based on measurement, or sampling from the probability distribution of the outcomes. 
We perform these simulations using the QASM simulator in Qiskit. 
We simulate the measurement procedure as it would happen on a noiseless quantum computer, and we also perform simulations using noise models based on the calibration parameters of a real quantum hardware. 
For our simulations we used the calibration of \textit{ibmq\_boeblingen}, that is a 20-qubit quantum device. 
This numerical study includes gate and measurement errors, statistical uncertainty, and decoherence. 
We repeat our measurement-based simulations with 2'048, 8'192, 32'768 and 100'000 shots for each step of the optimization.

For each set-up, 
we compare the two classes of ans\"{a}tze:
the first the is traditional VQE, where we only use the circuit ansatz, the second is  nu-VQE where we also apply the non-unitary operator. 
The optimization is performed using the Scipy Python package. For all the results presented in subsection \ref{Results_qutip}, the optimization is run 1000 times starting from different random initial angles and different random Jastrow parameters to avoid local minima, keeping the lowest energy to aim for the global minimum. 
For the results presented in section \ref{Results_noisy}, instead, the number of initial points is 100, since the simulations are computationally more expensive. 
For the classical optimization method we use in section \ref{Results_qutip} the BFGS optimizer, while in section \ref{Results_noisy} we used the Cobyla optimizer. 

\section{Results}
\label{Results}

\subsection{Results from exact noiseless simulations}
\label{Results_qutip}
\begin{figure*}[!htb]
\begin{tikzpicture}
\node[inner sep=0pt] (russel) at (-4,0)
    {\includegraphics[width=1.05\columnwidth]{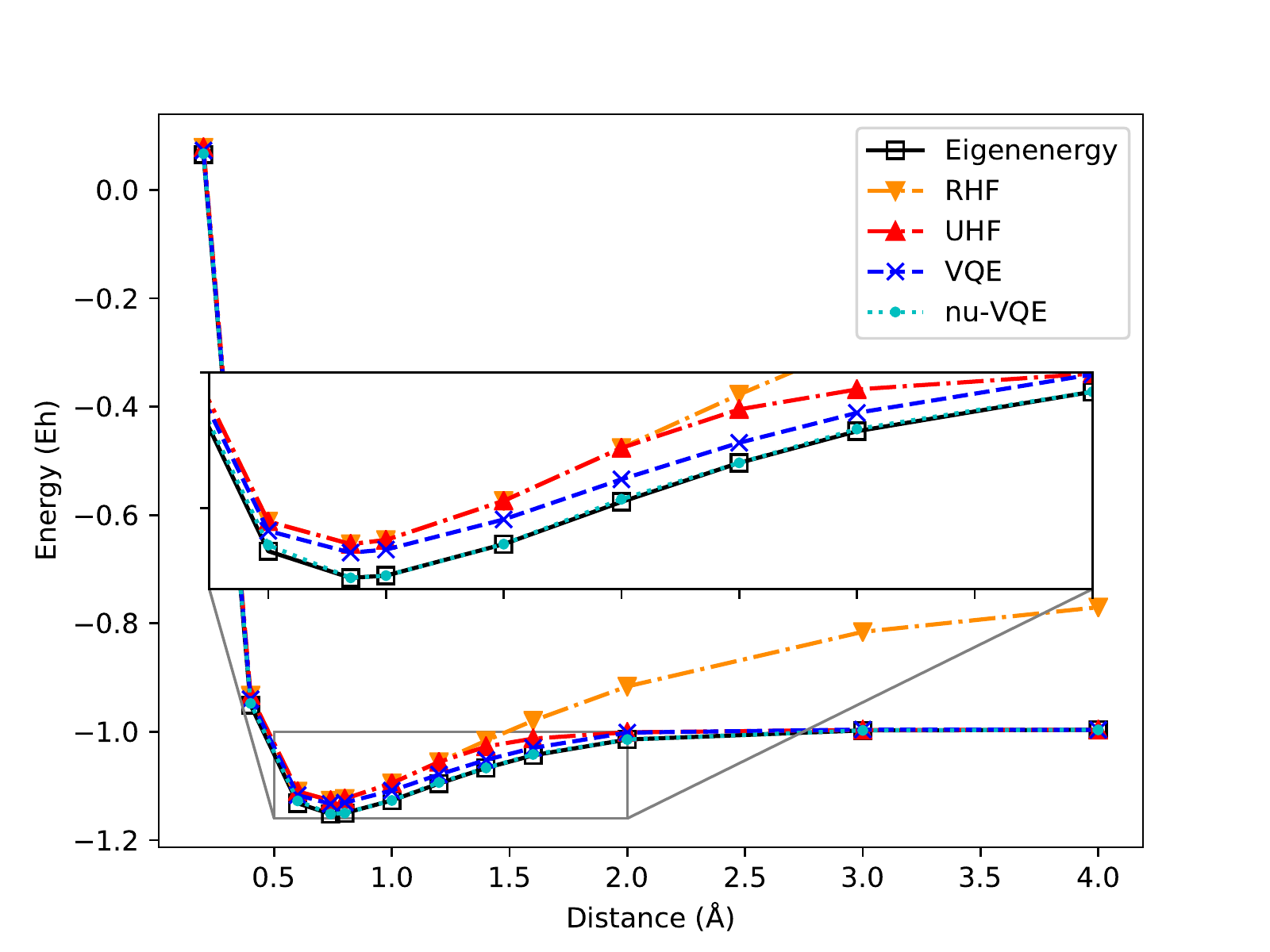}}; 
\node[text width=4cm, anchor=west, right] at (-8.6,2.5)
    {(a)};
\node[text width=4cm, anchor=west, right] at (0.15,2.5)
    {(b)};
    \node[inner sep=0pt] (russel) at (5.2,0)
    {\includegraphics[width=1.05\columnwidth]{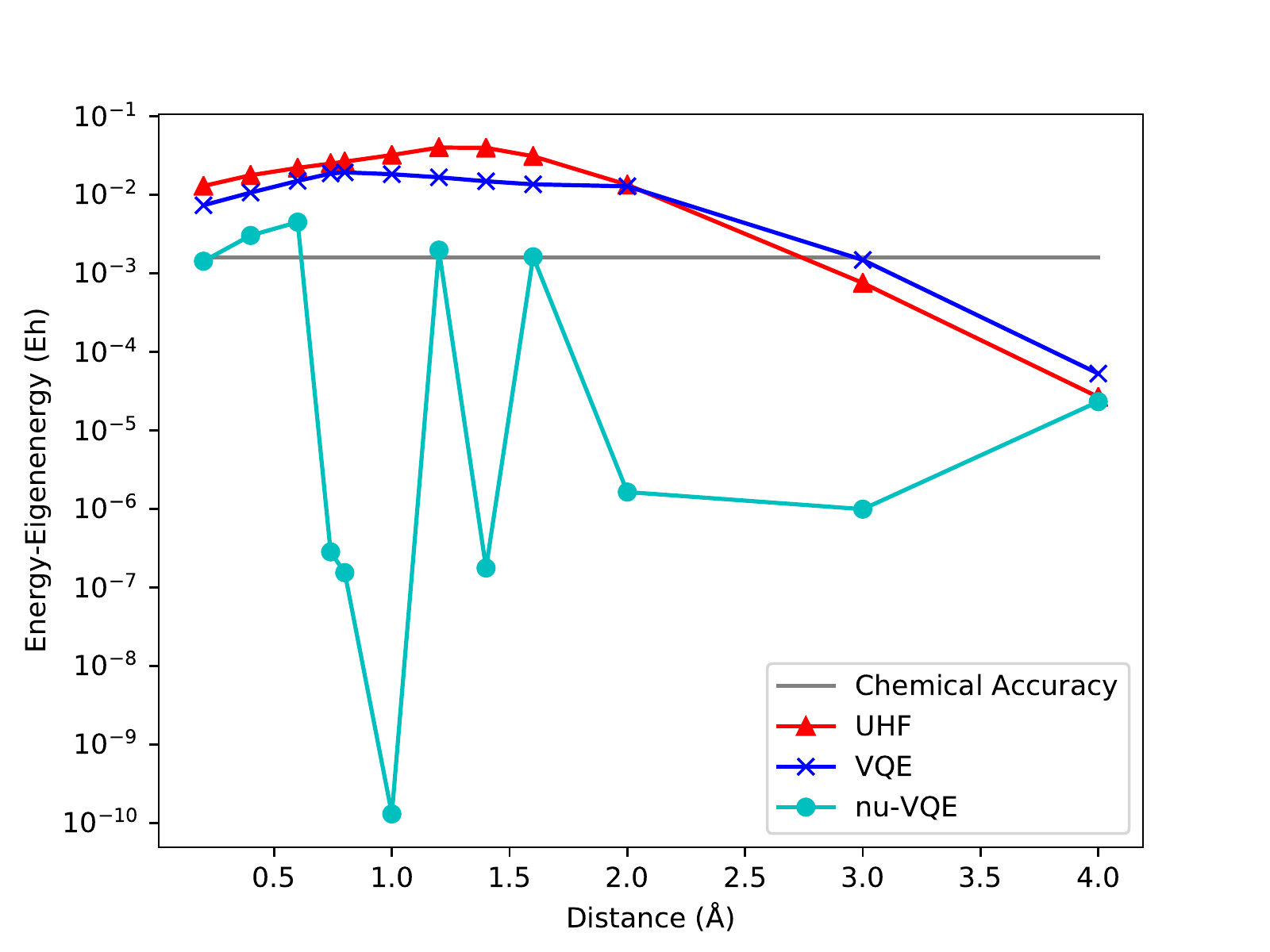}};
\end{tikzpicture}
\caption{$H_2$ dissociation curve. (a) Exact minimum eigenenergies (black), restricted Hartree-Fock energies (orange), unrestricted Hartree-Fock energies (red), VQE-optimized energies (blue), and nu-VQE-optimized energies (cyan) computed along the dissociation profile with the 6-31G basis set. With the Jordan-Wigner transformation the trial wavefunction is mapped in a 8 qubit register. 
(b) Corresponding energy errors computed as the difference between the VQE/nu-VQE energies and the exact lower eigenenergies of the qubit Hamiltonian at different bond lengths.
Color code as in panel (a).
The oscillations of the nu-VQE curve at small energy errors are due to convergence instabilities (note that we have 60 variational parameters in nu-VQE and only 24 in the regular VQE).
}
\label{fig:H2_curve}
\end{figure*}

We apply our methodology to a series of small molecules, namely molecular hydrogen ($H_2$), lithium hydride ($LiH$), and water ($H_2O$).
In Figure~\ref{fig:H2_curve} we report the dissociation energy curve for the $H_2$ molecule.
The electronic structure Hamiltonian is constructed using the 6-31G basis and mapped to a 8-qubits operator with the Jordan-Wigner transformation. 
The errors are defined as the difference between the minimized energy and the exact minimum eigenvalue of the Hamiltonian.
We observe that the errors obtained with the nu-VQE wavefunction are always at least one order of magnitude smaller than the corresponding results with the standard VQE, while in some cases the gain is even larger.


\begin{figure*}[!htb]
\begin{tikzpicture}
\node[inner sep=0pt] (russel) at (-4,0)
    {\includegraphics[width=1.05\columnwidth]{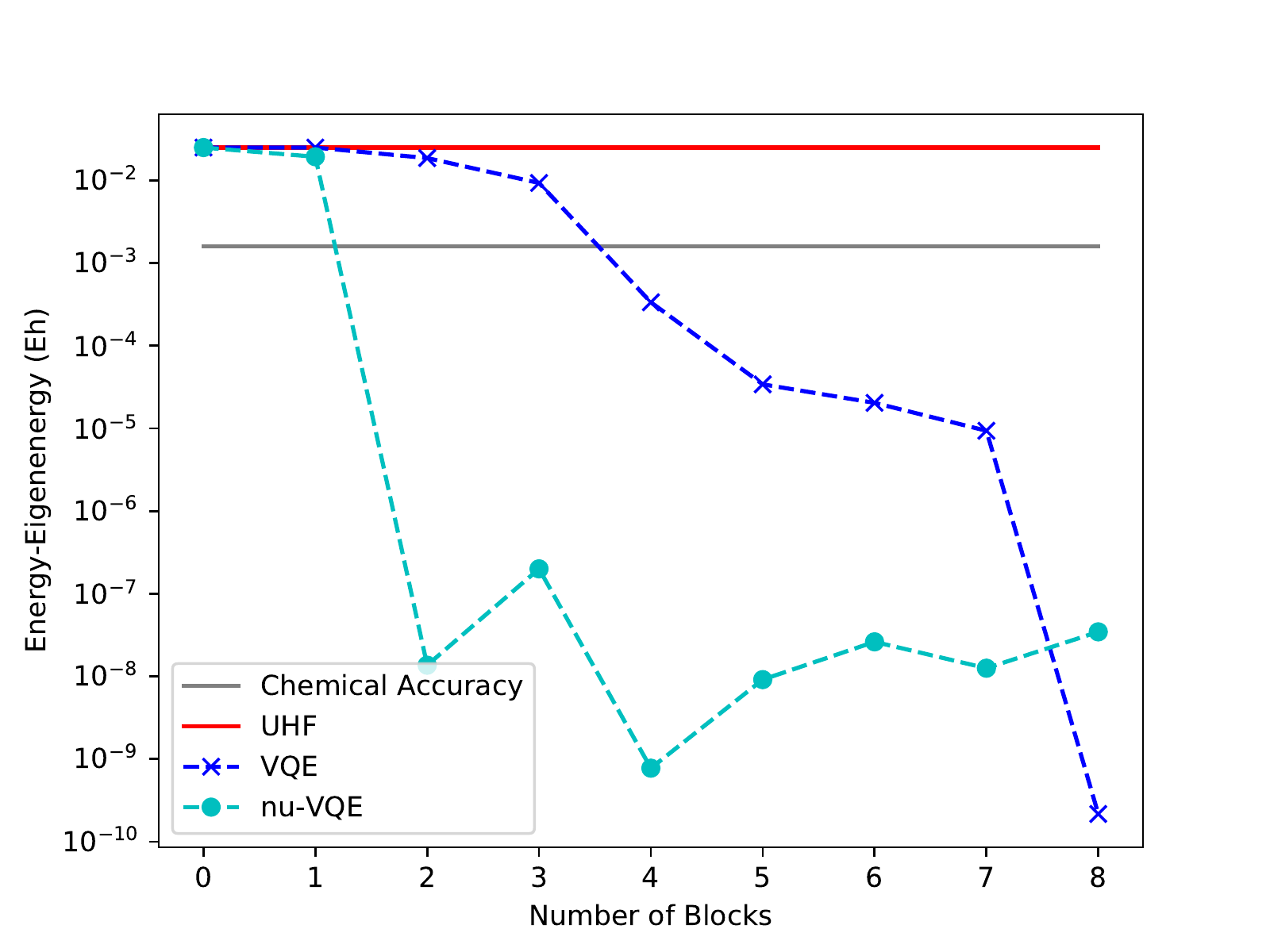}}; 
\node[text width=4cm, anchor=west, right] at (-8.6,2.5)
    {(a)};
\node[text width=4cm, anchor=west, right] at (0.15,2.5)
    {(b)};
    \node[inner sep=0pt] (russel) at (5.2,0)
    {\includegraphics[width=1.05\columnwidth]{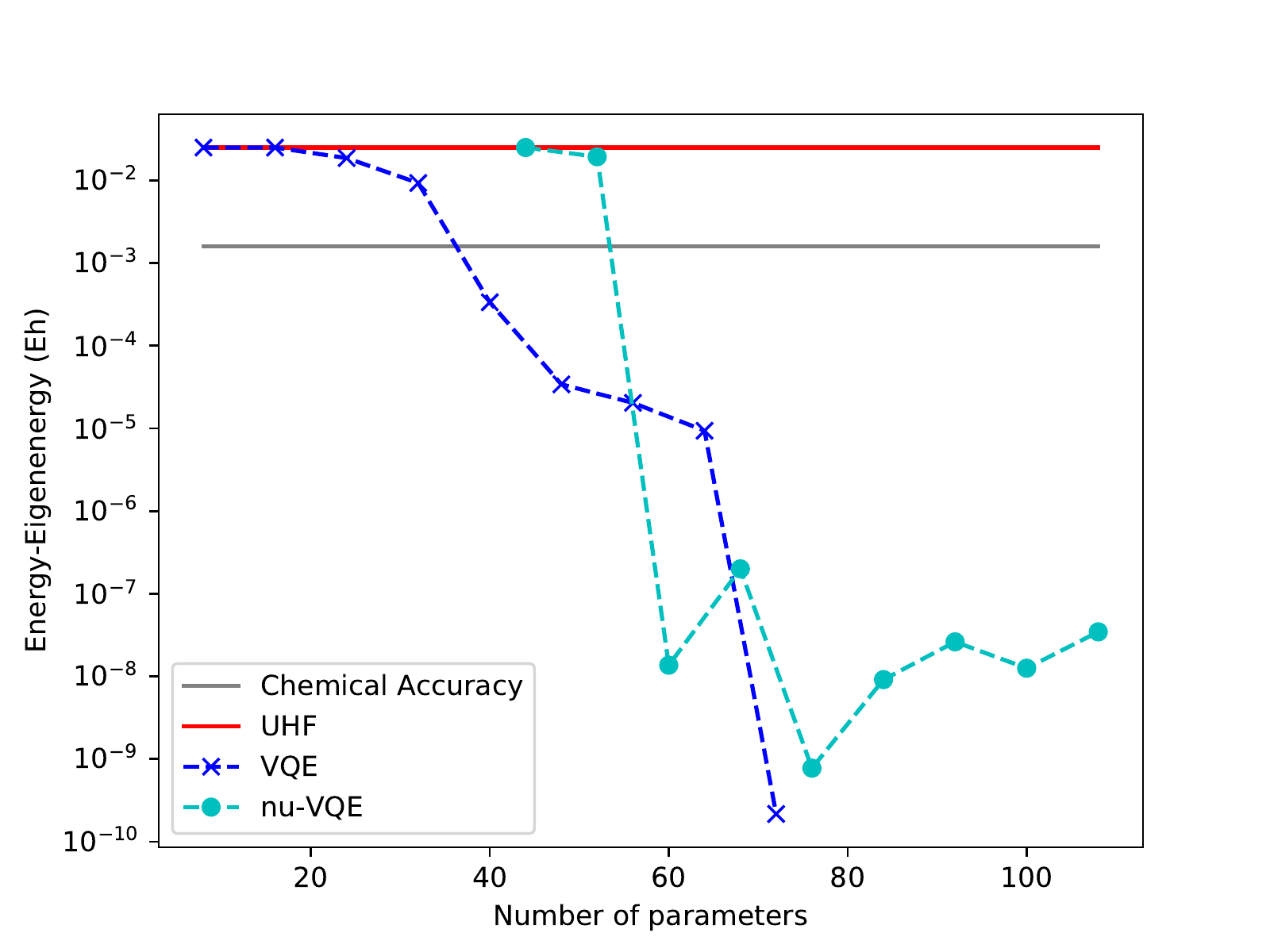}};
\end{tikzpicture}
\caption{$H_2$ simulations at the equilibrium distance of 0.74 \AA{} performed with the VQE and nu-VQE approaches
with the 6-31G basis set and the Jordan-Wigner mapping (8 qubits simulation). (a) Molecular energy errors as a function of the number of entangling blocks. Note that with nu-VQE we achieve chemical accuracy (gray line) with just 2 entangling blocks, reducing the number of gates required for convergence. (b) Same molecular energy errors as a function of the total number of variational parameters. 
}
\label{fig:vs_N_Blocks}
\end{figure*}

In Figure \ref{fig:vs_N_Blocks} we show that our algorithm can outperform the standard VQE, reaching chemical accuracy ($1 ~\text{kcal/mol}=1.6~\text{mEh}$) with about $50\%$ shorted circuits.
This is a consequence of the additional flexibility introduced by the variational Jastrow operator (see also Figure~\ref{fig:vs_N_Blocks}(b) for direct comparison in the number of parameters).



\begin{figure}[b]
\includegraphics[width=0.9\linewidth]{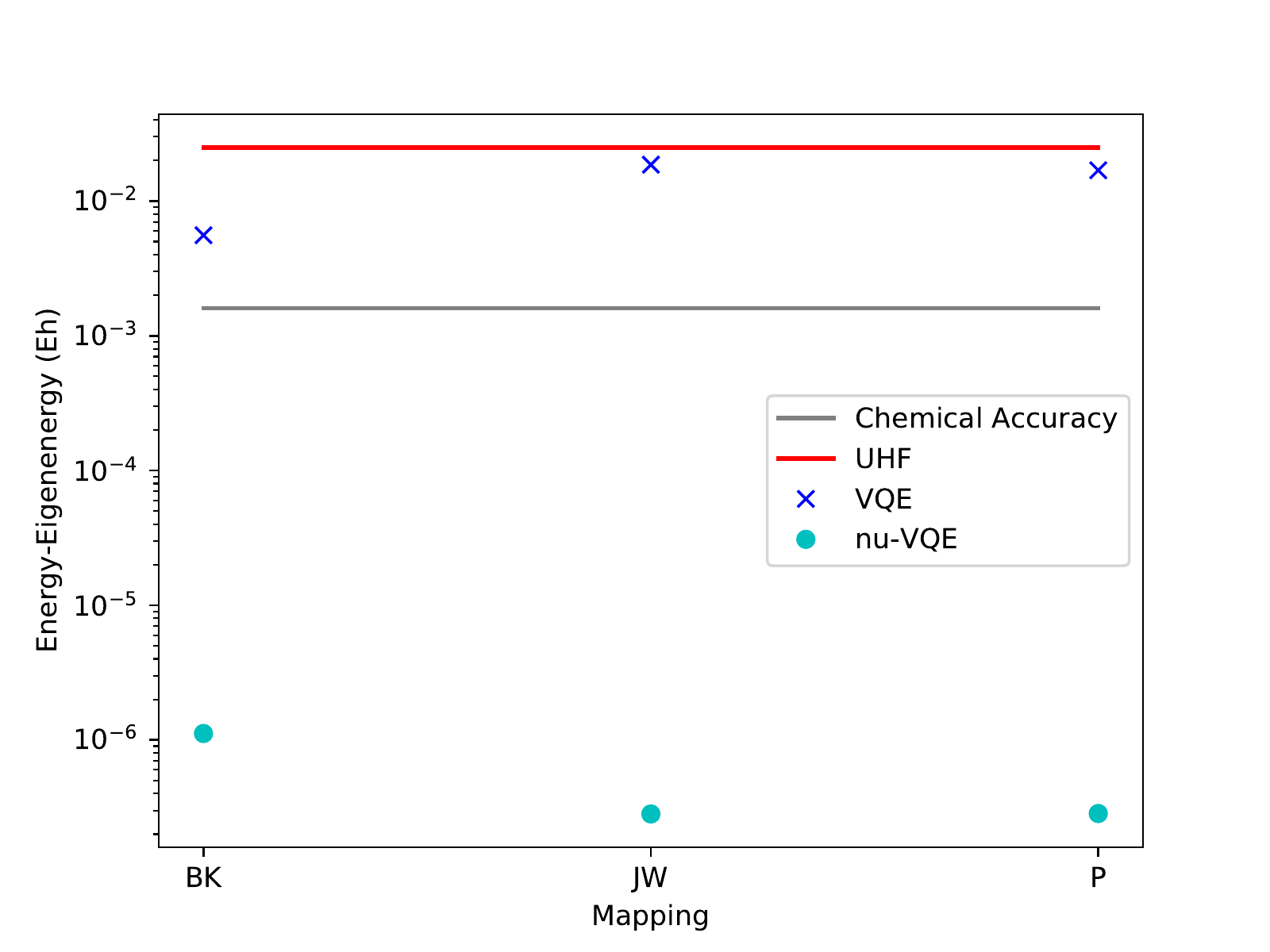}
\caption{
Deviations from the exact $H_2$ energy at the equilibrium distance 0.74 \AA{} obtained with the nu-VQE and the VQE approaches using 3 different fermion-to-qubit mapping schemes. As a reference, we also show the UHF (red curve) and the error corresponding to chemical accuracy (gray line).
}
\label{fig:vs_Mapping}
\end{figure}

The robustness of the nu-VQE method against different fermion-to-qubit mappings is tested in Figure \ref{fig:vs_Mapping}.
Interestingly, the method is successful for any chosen mapping, despite the fact that the physical interpretation of the Jastrow operator only holds in the case of the Jordan-Wigner mapping. 
Indeed, only in this case the combination of Eq.~\ref{eq:J1} and Eq.~\ref{eq:J2} corresponds to a qubit mapping of a sum of a fermionic one-particle density and a fermionic density-density Jastrow operators (notice that the Jordan-Wigner representation of the density operator acting on the fermionic mode $j$, can be rewritten as $\hat a_j^{\dagger} \hat a_j  = (Z_j+1)/2$).
This physical interpretation is lost when we use a different fermion-to-qubit mapping for the Hamiltonian, without changing Eq.~\ref{eq:J1} and Eq.~\ref{eq:J2} accordingly (as done in this work).
This is an indication that the nu-VQE method can also be effective for the calculation of the ground state energy of general Hamiltonians not related to electronic structure calculations.


\begin{figure}[ht]
\includegraphics[width=0.9\linewidth]{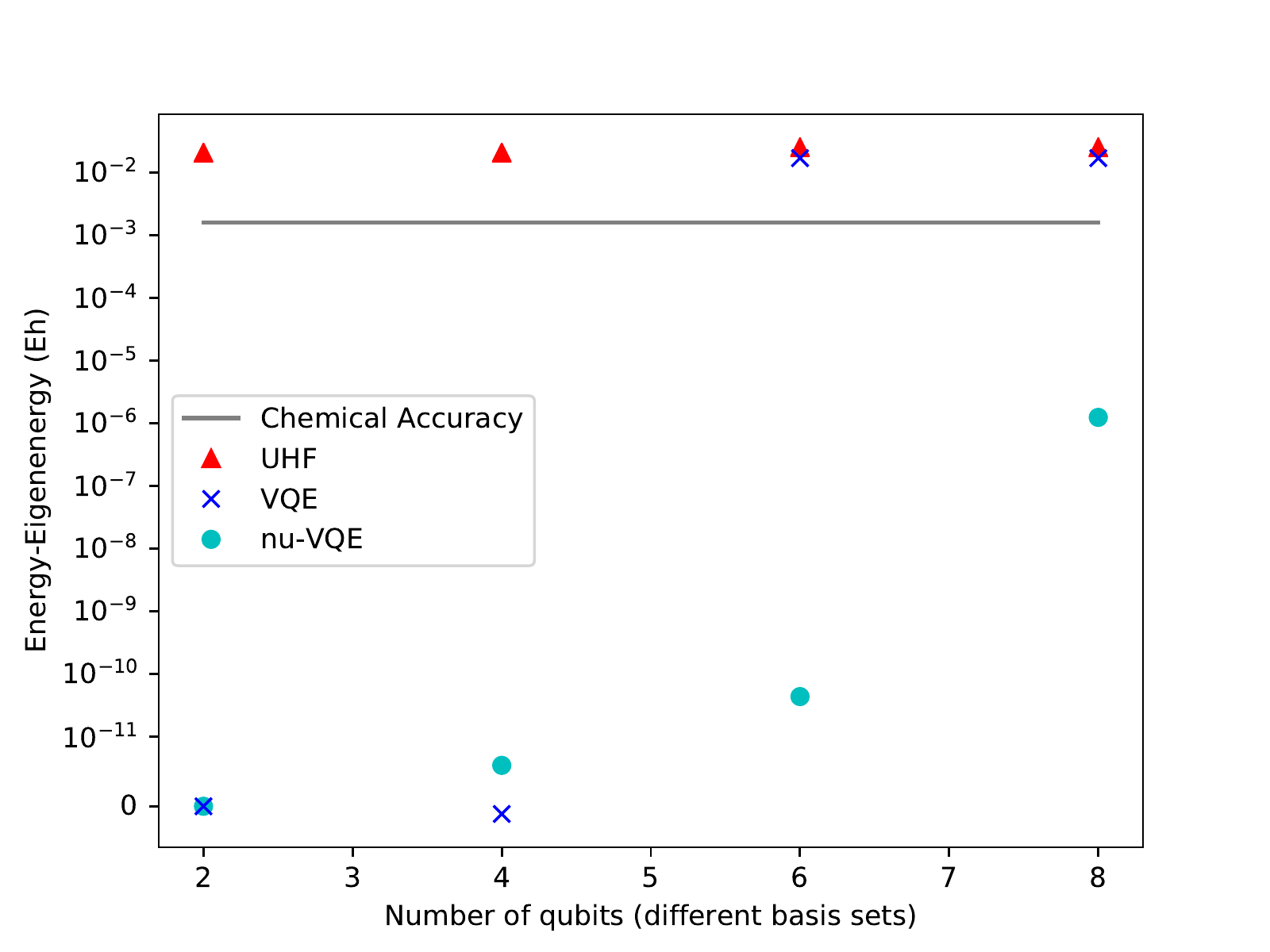}
\caption{
Deviations from the exact $H_2$ energy at the equilibrium distance 0.74 \AA{} obtained with the nu-VQE and the VQE approaches as a function of the number of qubits, which corresponds to 
the use of 2 different basis sets (STO-3G and 6-31G) with and without the application of the two-qubit reduction scheme.
}
\label{fig:vs_N_Qubits}
\end{figure}

\begin{figure}[ht]
\includegraphics[width=0.9\linewidth]{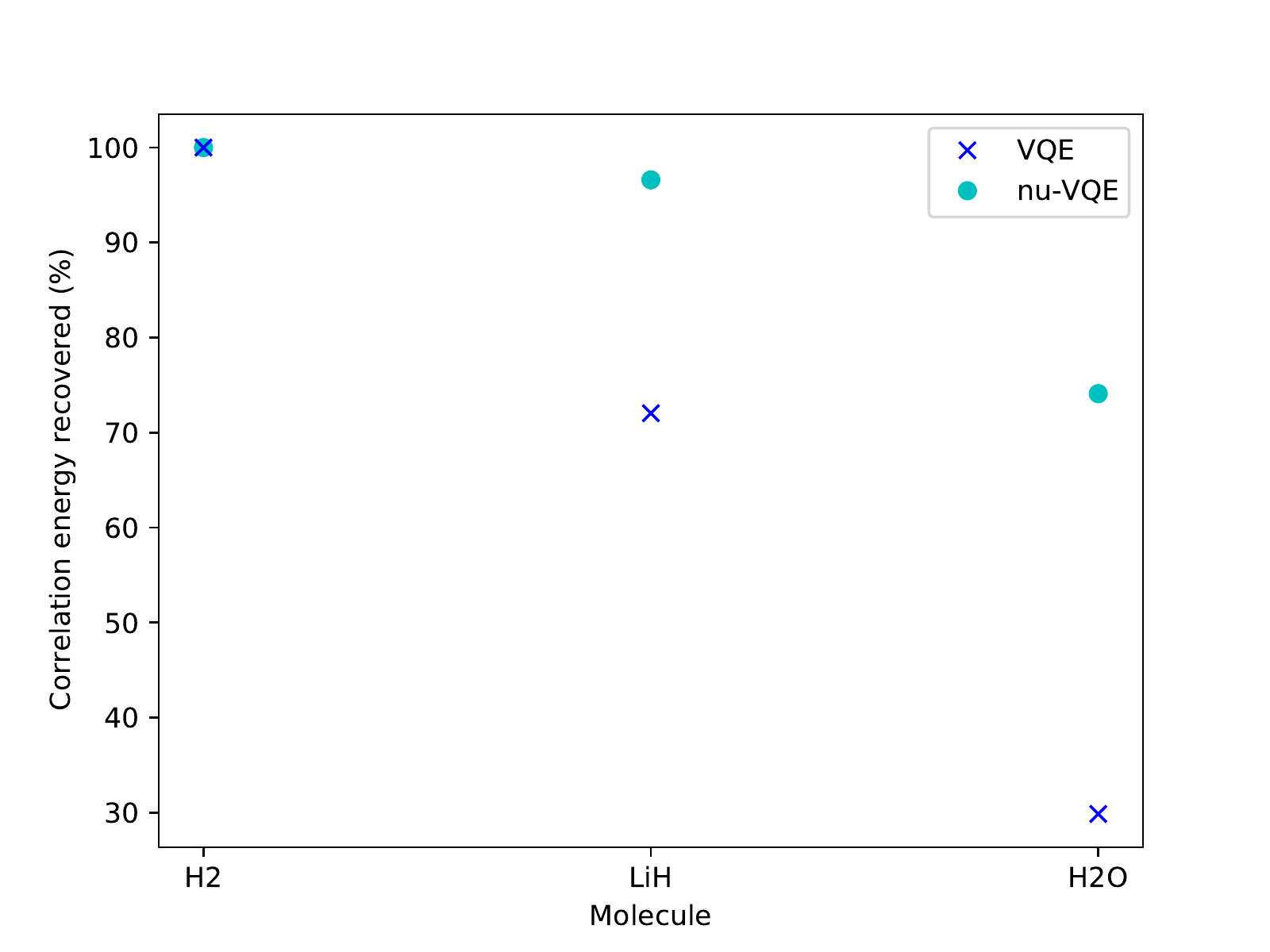}
\caption{
Amount of the total correlation energy recovered for the molecules $H_2$, $LiH$ and $H_2 O$ at the corresponding equilibrium geometries evaluated with the nu-VQE and the VQE approaches.
We used the STO-3G basis set with the parity fermion-to-qubit mapping and the two-qubit reduction scheme. The number of qubits for the 3 systems is 2, 10 and 12, respectively.}
\label{fig:vs_Molecule}
\end{figure}

Next, in Figure \ref{fig:vs_N_Qubits} we investigate the efficiency of the method as a function of the size of the qubit register.
To this end, we use the $H_2$ molecule and we vary the basis set from STO-3G (4 qubits) to 6-31G (8 qubits). 
In addition, we also exploit the possibility to use the two-qubit reduction scheme, which leads to 2 qubit and 6 qubit simulations with the STO-3G and the 6-31G basis sets, respectively.
However, it is important to note that by varying the atomic basis sets, we effectively generate every time a different qubit Hamiltonian.
For the larger number of qubits, the nu-VQE method is outperforming the VQE algorithm, whereas for smaller number of qubits (below 4) the results of the two methods are equivalent wit errors less than $10^{-10}$ Eh. 
In addition, nu-VQE reaches chemical accuracy for all examined systems, whereas VQE fails to converge to the correct ground state energy for the larger setups with 6 and 8 qubits. 
Additional insights can also be gained from the analysis of the simulations of the other molecular systems reported in  Figure~\ref{fig:vs_Molecule}.
Also in the cases of $LiH$ and $H_2O$ the amount of correlation energy recovered with nu-VQE is significantly higher than with standard VQE. 

Moreover, in view of future applications with noisy processors, we also investigated the stability of the nu-VQE energy expression in Eq.~\ref{en_eval}. 
A simple analysis reveals that in order to guarantee stability against noise, the denominator of Eq.~\ref{en_eval} needs to remain sizable all along the optimization path. 
In addition, another measure of the quality of the results consists in the evaluation of the Hartree-Fock contribution to the total wavefunction $|\Psi(\vec{\theta})\rangle$, which should remain dominant in most of the cases (with the exception for instance of multi-reference solutions and strongly correlated systems).
By imposing both these constraints we can avoid situations in which the uncertainty in the estimation of the denominator of Eq.~\ref{en_eval} becomes comparable to its absolute value, leading to important numerical instabilities.

\begin{figure*}[!htb]
\begin{tikzpicture}
\node[inner sep=0pt] (russel) at (-4,0)
    {\includegraphics[width=1.05\columnwidth]{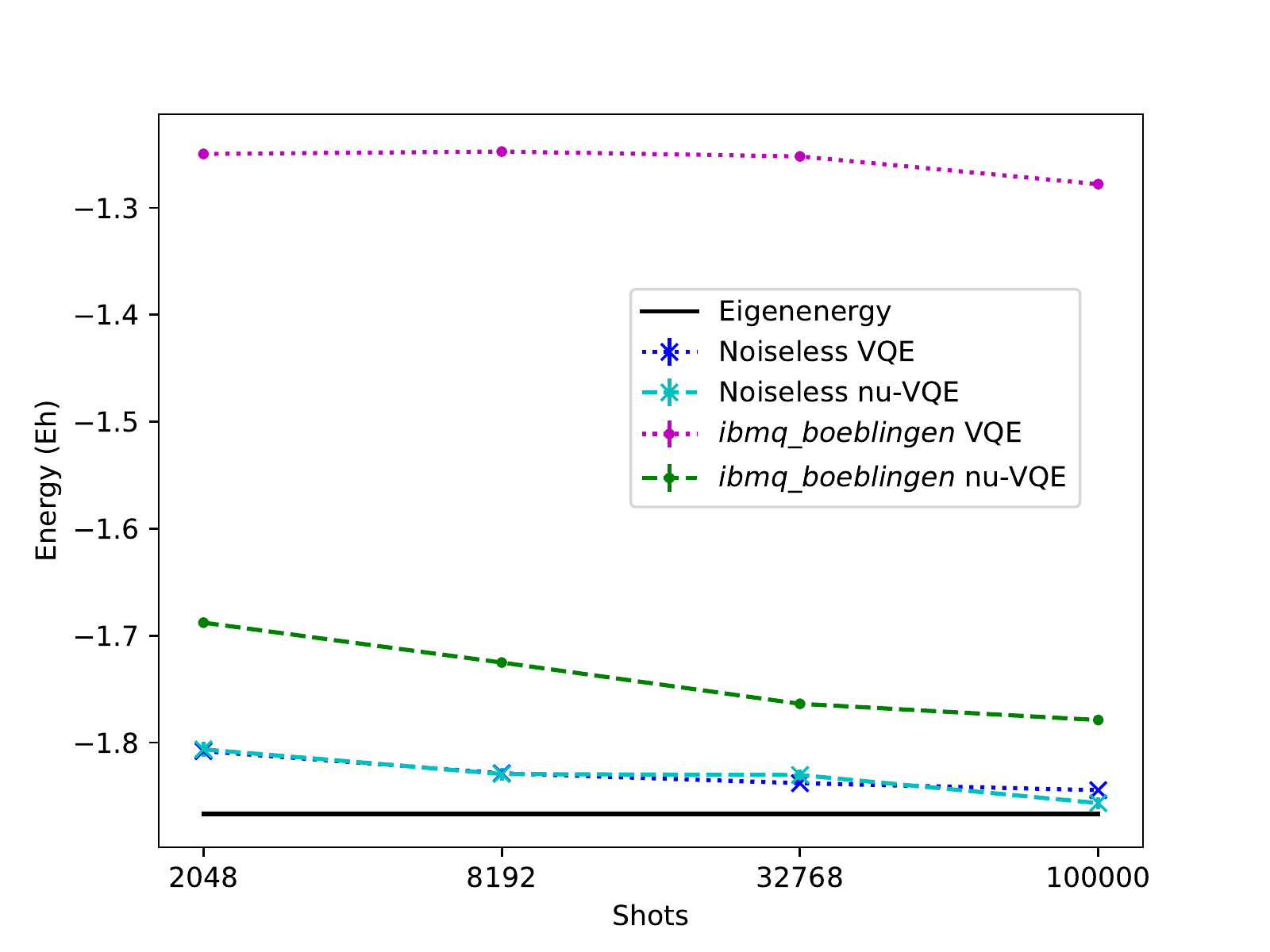}}; 
\node[text width=4cm, anchor=west, right] at (-8.6,2.5)
    {(a)};
\node[text width=4cm, anchor=west, right] at (0.15,2.5)
    {(b)};
    \node[inner sep=0pt] (russel) at (5.2,0)
    {\includegraphics[width=1.05\columnwidth]{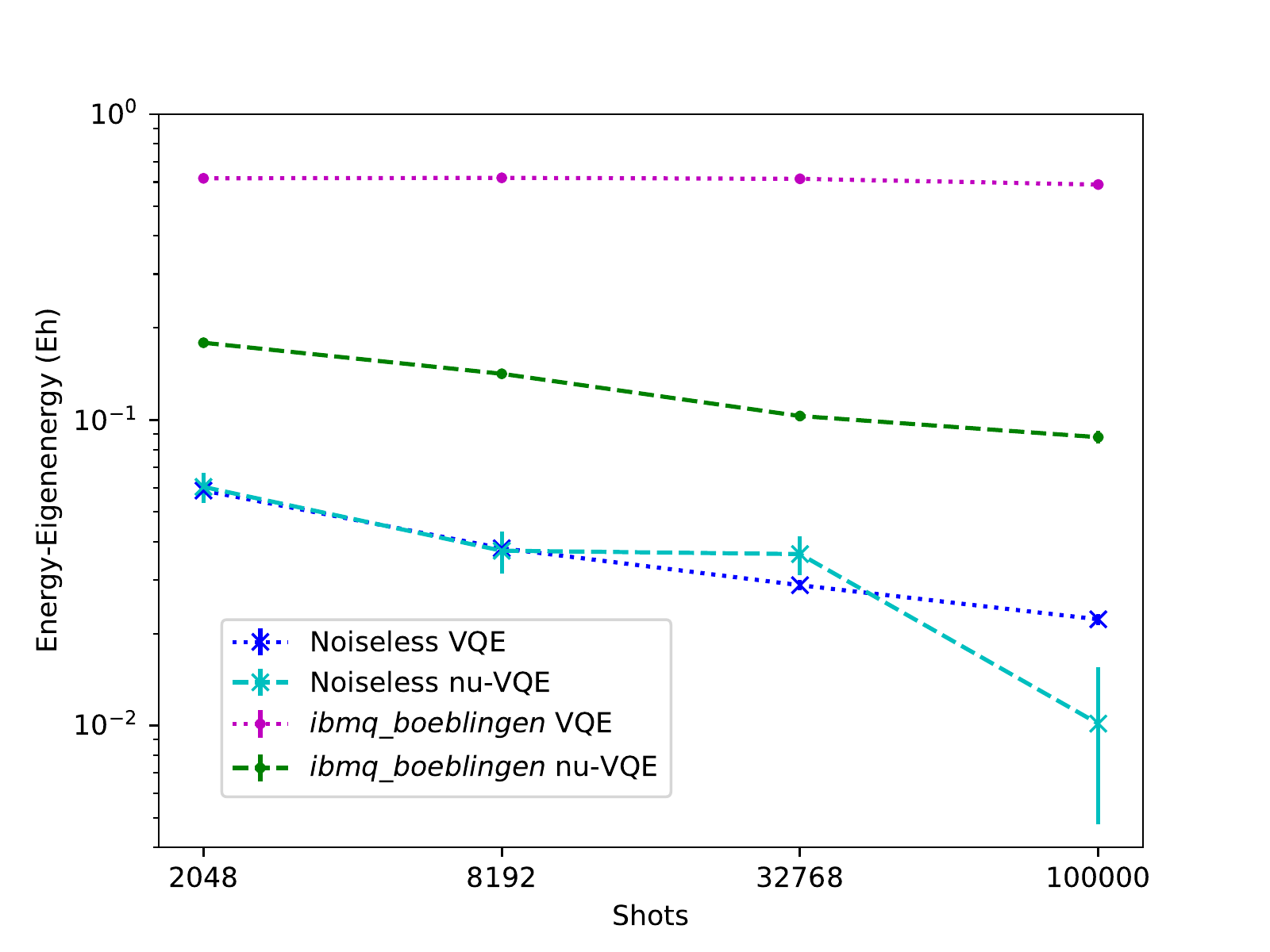}};
\end{tikzpicture}
\caption{(a) $H_2$ molecular energies obtained with the \textit{qasm\_simulator} at equilibrium distance 0.74 \AA{} using 2'048, 8'192, 32'768 and 100'000 shots for the evaluation of the expectation value of the Hamiltonian. Results are reported for the standard VQE and the Jastrow-enhanced nu-VQE approaches, with and without the presence of noise. In the case of the noisy simulations, we used the noise model and the parameters corresponding to the \textit{ibmq\_boeblingen} device.
(b) Corresponding relative errors.}
\label{fig:noisy}
\end{figure*}

\subsection{Results from measurement-based simulations} 
\label{Results_noisy}

For both the normal VQE and the new nu-VQE method, we studied the $H_2$ molecule at equilibrium distance with 6-31G basis, parity mapping and two-qubit reduction, leading to a 6 qubit Hamiltonian. When measuring the energy for each step of the optimization, we sample the distribution with the same number of shots. We ran the simulations using 2'048, 8'192, 32'768 and 100'000 shots for each step and for each simulation we measured the final energy with 100'000 shots using the optimized parameters.

\begin{figure*}[!htb]
\begin{tikzpicture}
\node[inner sep=0pt] (russel) at (-4,0)
    {\includegraphics[width=1.05\columnwidth]{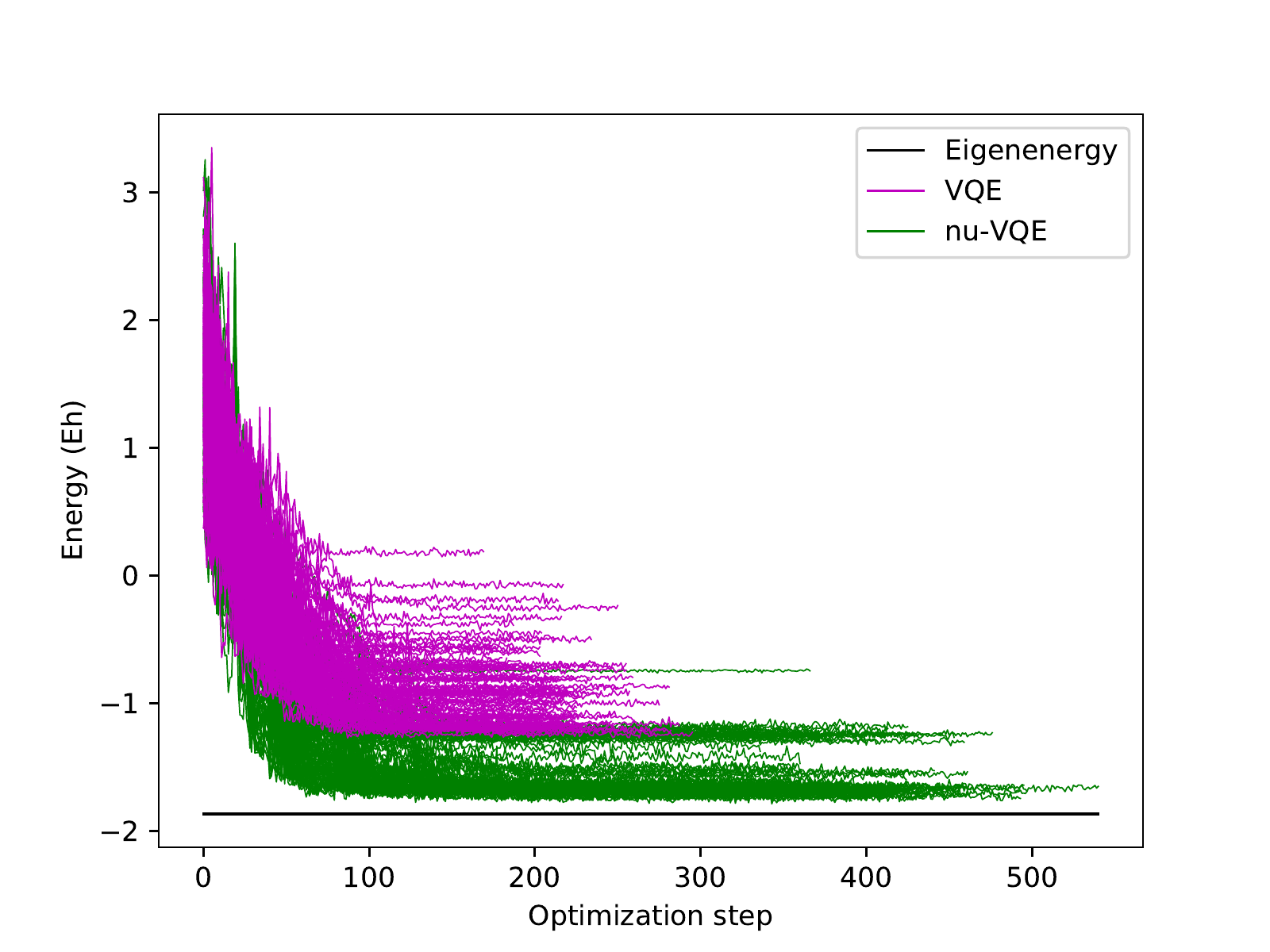}}; 
\node[text width=4cm, anchor=west, right] at (-8.6,2.5)
    {(a)};
\node[text width=4cm, anchor=west, right] at (0.15,2.5)
    {(b)};
    \node[inner sep=0pt] (russel) at (5.2,0)
    {\includegraphics[width=1.05\columnwidth]{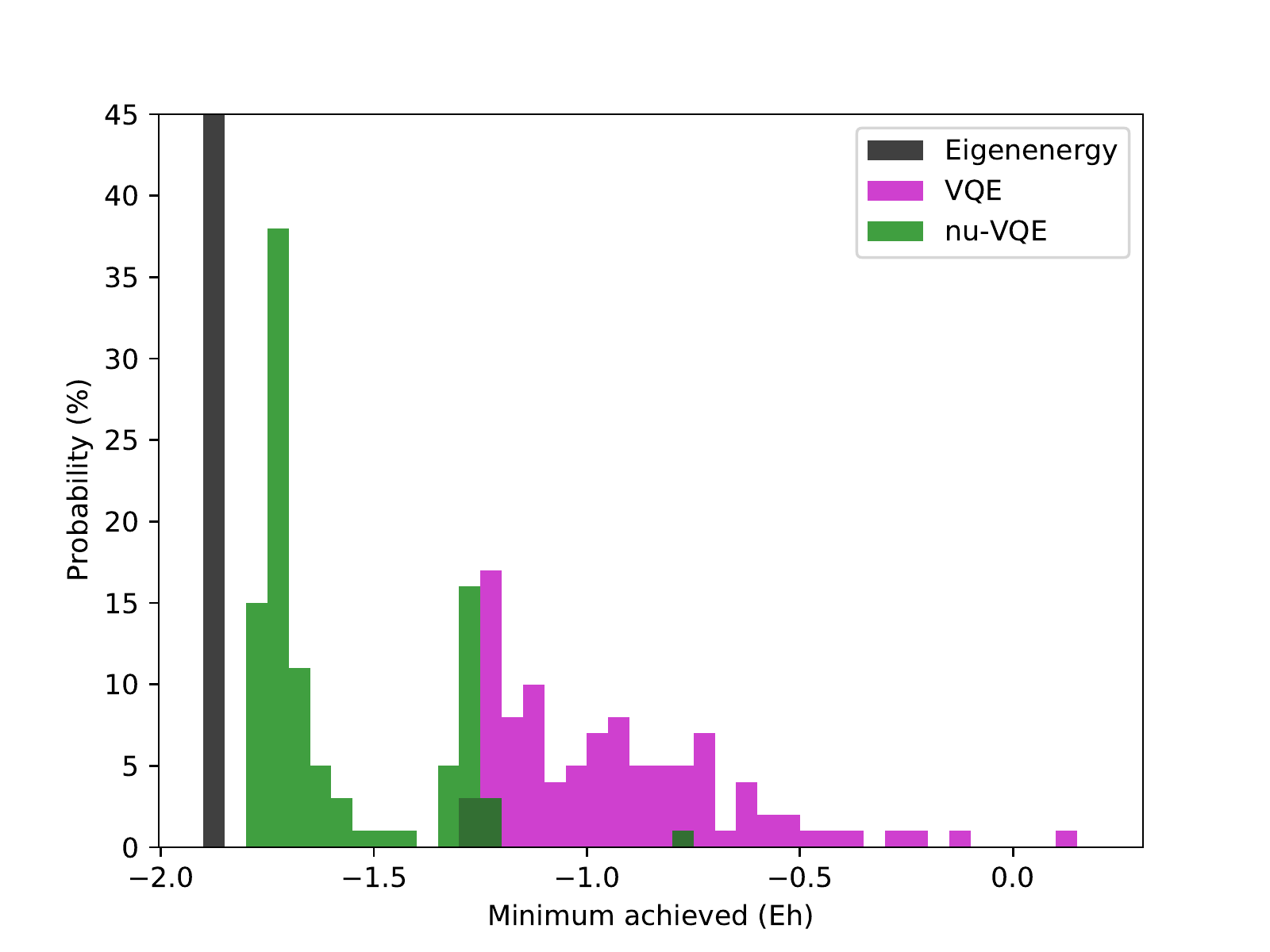}};
\end{tikzpicture}
\caption{(a) 
Evolution of the $H_2$ energies during the optimization with the VQE (without Jastrow) and nu-VQE (with Jastrow) algorithms. 
The simulations are performed using the 6 qubit Hamiltonian and the noise model of \textit{ibmq\_boeblingen} quantum computer. The lines correspond to 200 different initializations of the variational parameters.
(b) Histograms of the final VQE/nu-VQE energies for the 200 simulations. The width of each bin corresponds to 0.05 Eh.}
\label{fig:E_vs_iter}
\end{figure*}

In Figure \ref{fig:noisy}, we observe that when the simulations are run without any noise, both the nu-VQE and the VQE method display a similar error.
The error bars in these measurements, are almost invisible since they are all of the order of $10^{-3}$. They are simply the standard errors obtained from calculating the energy with 100'000 shots, and they do not take into account the large systematic error due to noise.
From panel (b) of Figure \ref{fig:noisy} we observe that the results obtained with the simulation of the nu-VQE algorithm on a noisy quantum computer are almost an order of magnitude more accurate than those obtained with the standard VQE. We can see that nu-VQE leads to an important error-mitigating effect, which is in fact the most important outcome of this method. 

In Figure~\ref{fig:E_vs_iter}, 
we compare different optimization runs started from different initial parameters using the standard VQE and the Jastrow-improved nu-VQE method. In the last case, the optimization was performed for 100 steps using the Jastrow factor and the remaining without). In all simulations, we used the \textit{ibm\_boeblingen} noise model and 8'192 shots for the evaluation of the expectation values. 
We notice (Figure~\ref{fig:E_vs_iter}(a)) that 
the number of optimization steps required by the nu-VQE method to reach convergence is about twice that of the normal VQE method. 
However, as shown in Figure~\ref{fig:E_vs_iter}(b), most of the optimizations performed with the nu-VQE approach converge to a final energy that is significantly closer to the exact eigenenergy than with standard VQE.


\section{Conclusions} 
\label{Conclusions}
We introduce a non-unitary Variational Quantum Eigensolver (nu-VQE) method, which makes use of a novel class of trial states for the variational quantum computation of molecular energies.
The new algorithm introduces additional variational parameters through the use of a non-unitary quantum operator on top of the traditional circuit ansatz. 
Our implementation extends the VQE to study non-normalized states, adding additional flexibility through a parametrized non-unitary operator. However, this comes with the additional cost of introducing more Pauli operators to measure.
Using an operator inspired by the so-called Jastrow factor of quantum Monte Carlo, we obtain effective wavefunctions that can approach the exact ground state with a shallower circuit. 
Although in this paper we focus only on physically inspired non-unitary operators, a different choice of the variational operators can be done. Alternatively, one may think about the possibility to increase rationally the complexity of the operator, similarly to what proposed, for instance, in the Adapt-VQE algorithm~\cite{Grimsley2019} or in the evolutionary optimization approach~\cite{rattew2019domainagnostic}.
Interestingly, our results show that the nu-VQE approach is equally effective also in the cases in which the Jastrow operator does not have any precise physical motivation, demonstrating a high level of flexibility and extensibility beyond the use in quantum chemistry applications. 
Moreover, the nu-VQE enables a strong noise mitigation effect~\cite{decoherence_mitigation}, improving by one order of magnitude the accuracy of the expectation values compared to the traditional VQE approach. 

\bibliographystyle{ieeetr} 
\bibliography{bibfile} 

\end{document}